\documentclass[12pt]{article}
\usepackage{amsmath,float,amssymb,amsthm,amsxtra,overpic,bbm,bm,epsfig,color}

\def\thefootnote{\fnsymbol{footnote}}

\def\tabnotefont{\fontsize{9}{10}\selectfont}%
\newenvironment{tabnote}{\par\tabnotefont}{\par}

\usepackage{slashed,stmaryrd}
\usepackage{color}

\usepackage[numbers]{natbib}
\usepackage[colorlinks=true,citecolor=blue,linkcolor=blue]{hyperref}
\usepackage{amsmath,amssymb}
\usepackage{graphicx,color}
\usepackage{sidecap}
\usepackage{footnote}
\makesavenoteenv{tabular}
\usepackage{xspace}

\begin{document}
\vspace{0.2cm}

\begin{center}
{\Large\bf A quantitative approach to select PMTs for large detectors}
\end{center}

\vspace{0.2cm}

\begin{center}
{\bf L.~J.~Wen $^{a}$}, \footnote{E-mail: wenlj@ihep.ac.cn}
\quad
{\bf M.~He $^{a}$},
\quad
{\bf Y.~F.~Wang $^{a}$},
\quad
{\bf J.~Cao $^{a}$},
\quad
{\bf S.~L.~Liu $^{a}$},
\quad
{\bf Y.~K.~Heng $^{a}$},
\quad
{\bf Z.~H.~Qin $^{a}$},
\\
{\small $^a$Institute of High Energy Physics, Chinese Academy of Sciences, Beijing 100049, China}
\end{center}

\vspace{1.5cm}

\begin{abstract}
Photomultiplier tubes (PMTs) are widely used in neutrino and other experiments for the detection of weak light. To date PMTs are the most sensitive single photon detector per unit area. In addition to the quantum efficiency for photon detection, there are a number of other specifications, such as rate and amplitude of after-pulses, dark noise rate, transit time spread, radioactive background of glass, peak-to-valley ratio, etc. All affect the photon detection and hence the physics goals. In addition, cost is another major factor for large experiments. It is important to know how to properly take into account all these parameters and choose the most appropriate PMTs. In this paper, we present an approach to quantify the impact of all parameters on the physics goals, including cost and risk. This method has been successfully used in the JUNO experiment. It can be applied to other experiments with large number of PMTs.
\end{abstract}

\begin{flushleft}
\hspace{0.9cm} Key words: neutrino detector, PMT selection, 20-inch PMT
\end{flushleft}

\def\thefootnote{\arabic{footnote}}
\setcounter{footnote}{0}

\newpage

\section{Introduction}
\label{sec:intro}

Since the first detection of neutrinos in 1956~\cite{Cowan56}, photomultiplier
tubes (PMTs) have been widely used in various neutrino detectors,
particularly the ones using liquid scintillator or water as target.
For neutrino detectors with PMTs, their performance drives the
accuracy and resolution of the energy, position or tracking,
and time measurements, and that is a critical factor to physics potential.
Typical PMT specifications include photon detection efficiency,
rate and amplitude of after-pulses, dark noise rate, transit time spread,
radioactivity of glass, peak-to-valley ratio, etc.

The quantum efficiency of the photocathode is the most
important factor that determines the PMT's ability
to detect photons. The photon detection efficiency (PDE) is
defined as $\frac{1}{S_{\rm n}}\int\limits_{\rm S_{pc}} \epsilon_{QE}\cdot\epsilon_{CE} dS$
in this paper. It represents the overall efficiency by averaging
the product of the quantum efficiency ($\epsilon_{QE}$) and
collection efficiency ($\epsilon_{CE}$) over different positions of
the photocathode. It should be pointed out that the actual
photocathode area ($S_{\rm pc}$) is usually smaller than
the nominal area ($S_{\rm n}$) given by the diameter of the PMT,
thus the fraction of effective photocathode area should be
considered when comparing different PMTs.
The thickness and composition of the photocathode vary over the cathode
and cause quantum efficiency non-uniformity. The collection efficiency
predominantly depends on the design of the focusing electrodes.

The transit time spread (TTS) affects the position
resolution of a point-like particle or the tracking resolution
of an energetic charged particle, particularly for large scale
detectors. The TTS is mainly affected by the shape of the PMT's glass
bulb, the design of the focusing electrodes and their location,
as well as the operating voltage. For large area PMTs,
simultaneous optimization of TTS and PDE was found to be difficult,
particularly if requiring the top and the equator of PMT's glass
bulb to have the same transit time. In some use case,
the equator region can be masked to guarantee good TTS throughout
the photocathode, resulting in less effective photocathode area.

Dark noise is mainly caused by the thermionic emission from the
photocathode. PMTs with high quantum efficiency or a large photocathode
intrinsically have larger probability of thermionic emission.
Random coincidences of PMT dark noise will trigger the detector and create
fake events, then the rate of dark noise affects the detector's energy
threshold and hence affects low energy neutrino studies.
The accidental coincidence of dark noise with physical signal degrades
the energy resolution. Operating at low temperature helps to
suppress the thermionic emission.

After-pulses are caused by the ion feedback from the ionization
of the residual gas or the amplification process.
The operating voltage affects the time distribution of after-pulses.
Pre-pulses are caused by a photon directly striking the collection
electrode without being reflected or absorbed by the glass or
photocathode. Both after-pulses and pre-pulses affect the
energy measurement.

The radioactivity of PMT's glass is an issue for low background
experiments. The glass and the other components of the assembly
inside the PMT should be screened before production till finding
the clean ones that satisfy the low radioactivity budget.

Neutrino oscillation studies require high precision measurements.
Future large detectors will be on the tens or
hundreds of kton scale. Tens of thousands of PMTs will be needed
if using liquid scintillator or water as target, and the
impact of PMT parameters on physics goals is more sensitive
comparing to a small scale detector. The PMT choices of some
recent neutrino detectors are listed in Table.~\ref{tab:exp}.
The small and median scale detectors such as
Daya Bay~\cite{DayaBay:2012aa}, Borexino~\cite{Alimonti:2000xc},
and SNO+~\cite{Boger:1999bb,Andringa:2015tza} choose 8-inch PMTs,
and some of them use light concentrators to enlarge the photon collection.
The large scale detectors such as
KamLAND~\cite{Eguchi:2002dm,Abe:2008aa},
Kamiokande-II~\cite{Hirata:1987hu},
Super-K~\cite{Fukuda:2002uc},
JUNO~\cite{An:2015jdp,Djurcic:2015vqa}
and Hyper-K 1TankHD~\cite{Abe:2018uyc}
unexceptionally use larger area PMTs.
Most of the detectors in Table.~\ref{tab:exp} choose
only one type of PMT. However, using a combination of different
types of PMTs may enhance the physics potential if their
performances are complementary.
Note that R\&D is still ongoing before the selection of the
PMTs for Hyper-K will be finalized. Herein, the selection of
JUNO PMTs is discussed.

Making the right choice of PMTs for large scale detectors is not
an easy task. For different physics topics, each PMT
characteristic affects the physics goals in a different way.
The impact of all PMT parameters should be taken into account,
and different types of PMTs need to be compared.
Cost is certainly a critical factor.
The risk regarding delivery capability and schedule also
need to be considered. We need a quantitative approach to choose
the most appropriate PMTs, considering the above factors and
various physics goals.

\begin{table}[!htb]
\begin{center}
\caption{PMT selection for some experiments.}
\label{tab:exp}
\begin{tabular}{cccc}
\hline
Experiment & Target mass & PMT type & PMT number \\ \hline

Daya Bay & 20 ton $^{\rm a}$ & 8-in. (R5912 $^{\rm b}$) & 192 \\
Borexino & 300 ton & 8-in. (EMI9351$^{\rm c}$) & 2,212 $^{\rm d}$ \\
SNO+ & 780 ton & 8-in. (R1408 $^{\rm b}$) & $\sim$9,300 $^{\rm e}$ \\
KamLAND & 1 kton & 20-in. (R3600 $^{\rm b}$) & 1,325 $^{\rm f}$ \\
                          &        & 17-in. (R7250 $^{\rm b}$) & 554 $^{\rm f}$ \\
Kamiokande-II & 3 kton & 20-in. (R1449 $^{\rm b}$) & 948\\
Super-K & 50 kton & 20-in. (R3600 $^{\rm b}$) & 11,146 \\
JUNO & 20 kton & 20-in. $^{\rm g}$ & $\sim$18,000 \\
Hyper-K 1TankHD & 260 kton & 20-in. $^{\rm h}$ & 40,000\\
\hline
\end{tabular}
\begin{tabnote}
  $^{\rm a}$ This is for one antineutrino detector module of the Daya Bay experiment.
  $^{\rm b}$ Produced by Hamamatsu Photonics K.\ K.
  $^{\rm c}$ Produced by Thorn EMI Electronics, which was merged by ET Enterprises Limited later.
  $^{\rm d}$ About 1,800 were equipped with light concentrators, and the remaining PMTs without light concentrators were used for distinguishing muon tracks in the buffer and point-like events in the scintillator.
  $^{\rm e}$ Each PMT was equipped with a 27 cm diameter concentrator.
  $^{\rm f}$ 554 are older Kamiokande 20-in PMTs and 1,325 are a newly developed, faster version masked to 17-in.
  $^{\rm g}$ It is a combination of MCP-PMT (GDG-6201) produced by North Night Vision Technology Co. Ltd. and dynode-PMT (R12860) produced by Hamamatsu.
  $^{\rm h}$ The selection of Hyper-K 20-in PMTs is still under R\&D stage, see Ref.~\cite{Abe:2018uyc}, thus the exact PMT type is not decided yet.
\end{tabnote}
\end{center}
\end{table}

The remaining part of this paper is organized as follows:
In Sec.~\ref{sec:futureDet}, we briefly introduce future large
detectors using PMTs. Two types of large area PMTs,
namely the MCP-PMTs and dynode-PMTs,
are discussed in Sec.~\ref{sec:pmtPar}.
In Sec.~\ref{sec:phys}, we describe a quantitative approach
to optimize the PMT choices for large neutrino detectors,
which takes into account the physics performance, cost and risk,
and the selection of JUNO PMTs is given as an example.
Finally, we summarize our study and discuss its prospects
in Sec.~\ref{sec:summary}.

\section{Future large detectors using PMTs}
\label{sec:futureDet}

\subsection{JUNO}

The JUNO experiment~\cite{An:2015jdp, Djurcic:2015vqa} is under
design and construction, and its central detector is an excellent
example of future large liquid scintillator detector.
The primary goal of JUNO is to determine the neutrino mass ordering.
In order to obtain an energy resolution better than
3\%/$\sqrt{E \text{(MeV)}}$,
maximizing the photon collection has been the most critical factor
driving the R\&D programs and the detector design. The JUNO central
detector consists of 20 kton purified liquid scintillator and two
independent PMT systems. The large PMT system has approximately
18,000 20-inch PMTs providing $\sim$75\% photocathode coverage,
and the average photon detection efficiency of PMTs is required
to be $>$27\% at 420 nm. The small PMT system has approximately
25,000 3-inch PMTs located in the gaps among the 20-inch PMTs,
providing an additional 3\% photocathode coverage and serving as
an independent calorimeter to calibrate the energy nonlinearity.
It also enhances the energy and track measurements for cosmic
muons and neutrino interactions. The water Cherenkov detector of
the JUNO veto system will be equipped with approximate
2,000 20-inch PMTs.

\subsection{Hyper-K}

Based on the experiences of Super-Kamiokande~\cite{Fukuda:2002uc},
the next generation water Cherenkov detector,
Hyper-Kamiokande~\cite{Abe:2018uyc,Abe:2011ts,Hyper-Kamiokande:2016dsw},
is under design. The Hyper-Kamiokande (Hyper-K) experiment continues
to use the water Cherenkov ring-imaging technique to detect neutrino
interactions and search for nucleon decays. The Hyper-K experiment has
planed two tanks and the total water mass will be 0.516 Mton.
The baseline design for the first tank, namely HK-1TankHD, is one
cylindrical vertical tank and the dimension of the water volume
is 74 m in diameter and 60 m in height. The inner active volume of
the tank, is a cylinder of 70.8 m in diameter and 54.8 m in height.
Hyper-K intends to keep 40\% photocathode coverage as SK-IV,
thus its inner detector is instrumented by approximate 40,000
inward-facing 20-inch PMTs. The outer volume, equipped with approximate
6,700 8-inch PMTs facing outwardly, acts as a veto for cosmic muons and
determines if a neutrino interaction occurring inside the inner
detector is fully contained.

\subsection{Neutrino telescopes}

The running neutrino telescopes in water and ice (like
ANTARES~\cite{Collaboration:2011nsa},
IceCube~\cite{Achterberg:2006md}) used the so-called optical module
which houses a single 10-inch PMT in a pressurized transparent vessel.
The recent KM3NeT~\cite{Bagley:2009wwa} experiment has been
developing the multi-PMT optical module concept,
which replaces the single large area PMT with 31 3-inch PMTs.
Such design has a better granularity with an accurate photon counting,
and potentially fast timing for multi event reconstruction.
It is more robust against the Earth's magnetic field and the loss of
one single PMT. In addition, thanks to the external pressurized
vessel, the risk of chain implosion of PMTs is largely suppressed.
Thus Hyper-K takes this concept as one of its photosensor
alternatives, and has been developing multi-PMT module with
similar dimension as KM3NeT but using UV transparent acrylic to
make the pressure vessel.

\section{Characteristics of large area PMTs}
\label{sec:pmtPar}

The 20-inch PMTs were motivated by the Kamiokande experiment and
invented in the 1980s~\cite{Kume:1983hs}. The quality was improved
with a ``Venetian-blind" type dynode (Hamamatsu R3600) used in
the Super-Kamiokande detector~\cite{Suzuki:1992as}. A version masked
to 17 inches with a ``box-and-line" type dynode (Hamamatsu R7250) was
developed for the KamLAND detector~\cite{Eguchi:2002dm},
and it provided faster timing for vertex reconstruction.
The above three experiments have demonstrated that large area PMTs
have the best performance-to-price ratio for large scale detectors.
However, due to technical difficulties of fabricating 20-inch PMTs,
very limited options were available and there was little space for optimization in the past.

Since JUNO and Hyper-K were proposed in late 2000s, the baseline
choice of their photosensors has been using 20-inch PMTs.
The available 20-inch PMT at that time was still the Hamamatsu R3600 model,
with an average quantum efficiency of $\sim\,$22\%
and collection efficiency of 70\% at $\sim$390 nm~\cite{Suzuki:1992as}.
It satisfied the early minimum requirement of Hyper-K inner detector
photosensors~\cite{Hyper-Kamiokande:2016dsw}.
Nevertheless, the PDE of R3600 model is only about half of the required
PDE for JUNO. Thus a completely novel design using a Micro-channel
Plate (MCP) in place of a dynode to amplify photoelectrons was
proposed~\cite{Wang:2012rt,NeuTel2017:Wang}. The original design of large area
MCP-PMT had the transmission photocathode coated on the front
hemisphere and the reflection photocathode coated on the rear
hemisphere to form nearly 4$\pi$ coverage to enhance the PDE. In 2015,
the quantum efficiency of transmission photocathode reached 30\% with
improved transfer bialkali-photocathode technology~\cite{Qian-NNN16},
and the reflection photocathode on the rear hemisphere was abandoned,
which improved the transit time spread and suppressed the dark noise. Furthermore,
several improvements were made to reach almost 100\% collection efficiency,
by optimizing the diameter of micro-channel plate,
the size and inclined angle of the micro-channel, the open area ratio,
and particularly using atomic layer deposition technology~\cite{ALD}.
As an outcome of maximizing the collection efficiency,
it has relative larger transit time spread than the Hamamatsu dynode PMT.

During the MCP-PMT development process, Hamamatsu improved its
dynode 20-inch PMTs with higher quantum efficiency photocathode and better
box-and-line dynode design~\cite{Nakayama-NNN14}, and the new type
R12860-HQE became available. Given the recent developments on large
area PMTs, in the most recent Hyper-K design report~\cite{Abe:2018uyc}
the minimum requirement on PDE was increased to 26\% around 400 nm.
Another new type of 20-inch PMT, the so-called hybrid photodetector,
was developed to obtain better timing and charge resolution.
It uses an avalanche photodiode (APD) to replace the dynode for the
amplification of photoelectrons. To guarantee the collection
efficiency on the small area of the APD, $\sim$8 kV high voltage
need be applied between the APD and photocathode, thus its safe use
in water over many years needs to be demonstrated.

By 2015, the 20-inch MCP-PMTs delivered by NNVT company (North Night
Vision Technology Co. Ltd.) had comparable performance with respect
to the 20-inch dynode-PMTs from Hamamatsu. Their typical
specifications are summarized in Table.~\ref{tab:pmtPar}.

\begin{table}[!htb]
\begin{center}
\caption{The typical specifications of the 20-inch MCP-PMT and dynode-PMT in 2015. The typical values, as well as the lower or upper limits, are listed.}
\label{tab:pmtPar}
\begin{tabular}{c|c|c}
\hline
Characteristics & MCP-PMT & Dynode-PMT (R12860)\\
\hline
Detection Efficiency$^{\rm a}$ [\%] & 27, $>$24 & 27, $>$24 \\
Dark noise rate$^{\rm b}$ [kHz] & 20, $<$30  & 10, $<$50\\
                             & $^{238}$U : $<$50 & $^{238}$U : $<$400 \\
Radioactivity of glass [ppb] & $^{232}$Th : $<$50 &  $^{232}$Th : $<$400\\
                             & $^{40}$K : $<$20 &  $^{40}$K : $<$40 \\
Transit Time Spread$^{\rm c}$ (FWHM) [ns] & 12, $<$15 & 2.7, $<$3.5 \\
Pre-pulsing/After-pulsing [\%] & $<$1 / $<$2 & $<$1.5 / $<$15 \\
Rise time/Fall time$^{\rm d}$ [ns] & 2 / 12 & 5 / 9\\
Peak-to-Valley ratio & 3.5, $>$2.8 & 3, $>$2.5\\
\hline
\end{tabular}
\begin{tabnote}
  $^{\rm a}$ The quoted detection efficiency refers to 420 nm photons.
  $^{\rm b}$ Measured with a threshold of 1/4 p.e.
  $^{\rm c}$ Measured on the top point of PMT.
  $^{\rm d}$ The quoted rise and fall time refers to single photoelectron waveforms.
\end{tabnote}
\end{center}
\end{table}

The 20-inch PMT's glass bulb is typically made by borosilicate glass,
and it is non-trivial to control the radio-purity. The raw material
components, such as silica sand, borax and boracic acid, aluminum
hydroxide and industrial salt, should be screened until finding the
clean ones that satisfy the low radioactivity budget.
For the MCP-PMT's glass bulb, the projected radioactivity before mass
production is listed in Table.~\ref{tab:pmtPar}, by summing up the
selected materials according to their weight fractions. During the
production cautious procedures were established and executed to
control the radioactivity~\cite{Zhang:2017ocm}, and the final
radioactivity slightly differ from the original projection.
Unfortunately it was impractical to follow the same approach to
control the glass radioactivity for Hamamatsu products.
Table.~\ref{tab:pmtPar} also lists the projected radioactivity
based on a few measured glass samples of Hamamatsu R12860.

The pre-pulsing ratio of MCP-PMTs is negligible because the surface
of micro-channel plate has very low photoelectric conversion probability,
whereas the dynode PMTs have a typical pre-pulsing peak
about 10-90 ns before the main peak. The after-pulsing probability
for MCP-PMTs is also lower, most likely due to the different structures
between micro-channel plate and dynode and less residual
gas which results in smaller ion feedback. As for other characteristics,
both products guarantee good peak-to-valley ratio. The MCP-PMT has higher
dark noise rate than the dynode PMT.
The single photon electron waveforms of the MCP-PMTs have much faster
rise time due to the fast amplification inside the micro-channels.

\section{Optimize PMT selection for large detectors}
\label{sec:phys}

\begin{figure}[!htb]
\centering
\includegraphics[width=0.6\textwidth]{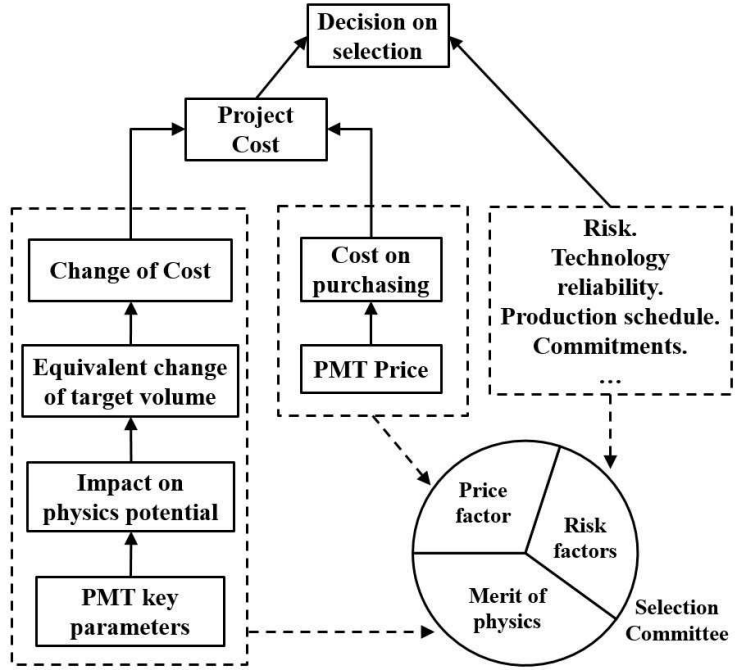}
\caption{Diagram of the quantitative approach to select PMTs in JUNO.}
\label{fig:diag}
\end{figure}

For future large scale detectors, establishing a science-driven
approach for the procurement of tens of thousands PMTs is highly
desired. The selection should be based on the physics performance,
price and risk. Below we propose a quantitative approach:
\begin{itemize}
  \item Quantify the impact of each key parameter of PMT on physics goals;
  \item Convert the changes of physics performance into equivalent monetary cost, so that the impact on physics and real cost on purchasing PMTs can be normalized to the changes of overall project cost.
  \item Evaluate the risk, technology reliability, production schedule, service commitments, etc, and combine with the above outcome to make a decision.
\end{itemize}
Such approach is used in 2015 for the selection of JUNO 20-inch PMTs,
and the diagram shown in Fig.~\ref{fig:diag} illustrates the concepts.
The PMT selection is made by a selection committee
based on three requirements:
physics performance, cost, and risk factors.
Different weights can be assigned to these different requirements
and the physics performance is the most critical one.
Taken JUNO as an example, the details of this approach are discussed
below, particularly the merit of physics.

\subsection{An example: JUNO 20-inch PMTs selection}

\subsubsection{How photo-sensors drive the sensitivity}

In JUNO, the neutrino mass ordering is determined via precise
spectral measurements of reactor antineutrino oscillations.
Energy resolution is the most crucial factor to probe the
interference effect of two fast oscillation modes.
A conventional parametrization of energy resolution can be written as
$\sigma_E/E=\sqrt{a^2/E + b^2 + c^2/E^2}$,
where it consists of the stochastic term $a$, the constant term
$b$ and the noise term $c$. These three terms play different roles in
affecting the sensitivity to neutrino mass ordering,
thus an effective description is
$\sqrt{(a^2 + (1.6b)^2 + (c/1.6)^2)/E}$~\cite{An:2015jdp}.
The impact of the main PMT characteristics on energy resolution is
summarized below.
\begin{itemize}
  \item The PDE drives the stochastic term $a$, which is the most
  critical parameter for determining the neutrino mass ordering.
  The minimum required PDE for the JUNO 20-inch PMTs is 27\%.

  \item The dark noise contributes predominantly to the noise
  term $c$. Occasionally dark noise coincides with a physical event and degrades the energy resolution.

  \item The constant term $b$ can be caused by the Cherenkov process,
  quenching effects, instability of detector response,
  residual energy errors after non-uniformity correction, and
  non-linearity of the electronics.
  For a large liquid scintillator detector like JUNO, the non-zero
  transit time spread degrades the vertex resolution~\cite{Liu:2018fpq}
  and consequently propagates into the energy error through non-uniformity correction, and eventually contributes to the constant term of the energy resolution.

  \item The pre-pulses arrive a short time prior to the main pulse,
  and they can directly overlap with true photons.
  After-pulses have instead a wider time distribution.
  Because the electron anti-neutrinos are detected via the inverse beta
  decay reaction, that consists of a prompt positron signal and a
  delayed neutron-capture signal, the after-pulsing of the positron
  signal may fall into the signal window of the neutron signal and shift
  the neutron energy, or it could overlap with positron's own photons
  and shift the positron energy.
  Such effect will then hurt the accuracy of energy measurement and eventually degrade the sensitivity to neutrino mass ordering.
\end{itemize}

The total weight of the PMT glass in JUNO is $\sim\,$140 tons,
then the radioactivity of PMT's glass bulb would lead to non-zero
background even with a water buffer between the liquid scintillator target
and glass. A higher radioactivity requires tighter fiducial volume if
keeping the same signal to background ratio, resulting in a degradation
of sensitivity.

\subsubsection{PMT performance merit}
\label{sec:physMerit}

\begin{figure}[!htb]
\centering
\includegraphics[width=0.5\textwidth]{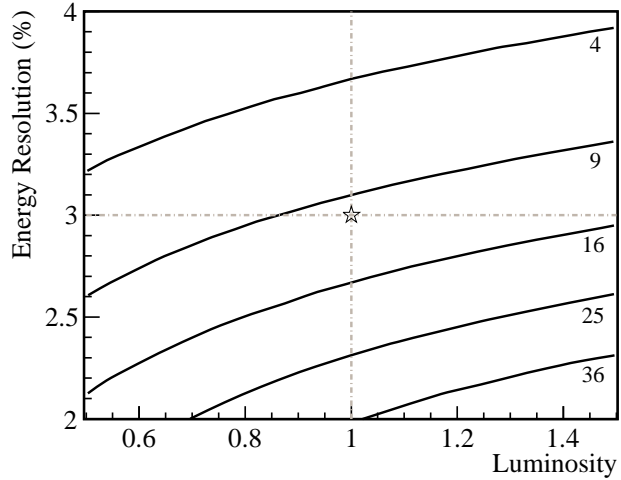}
\caption{This plot is revised from Figure 13 in~\cite{An:2015jdp}. It shows the $\Delta\chi^2_{MH}$ contour as the function of the event statistics (luminosity) and the energy resolution, where the open star marker stands for the 3\%/$\sqrt{E \text{(MeV)}}$ energy resolution and the nominal running of six years with 20 kt target mass, 36 GW$_{\rm th}$ reactor power and 80\% efficiency.} \label{fig:EresLu}
\end{figure}

As discussed above, the variation of each PMT characteristic affects
differently the sensitivity to neutrino mass ordering,
by changing either the energy resolution or the background.
In Fig.~\ref{fig:EresLu} (taken from Figure 13 in~\cite{An:2015jdp}),
it shows the sensitivity ($\Delta\chi^2_{MH}$) contour as
a function of the event statistics and the energy resolution.
Thus, if setting a precondition of keeping the same projected sensitivity
to neutrino mass ordering in six years, the change of energy resolution or
background can be scaled to a corresponding change of the target mass.
The increase or reduction in the total cost due to the change of the
target mass, as well as the consequent cost change of acrylic sphere,
number of PMTs, stainless steel structure and water pool excavation
can be estimated according to their construction costs,
and it quantifies the impact of the PMT performance.

For parameters such as PDE, dark noise rate or transit time spread,
any deviations from the reference values were converted into changes
of energy resolution by Monte Carlo studies. Their performance merit
was further evaluated following the above method and parameterized
in Eq.~\ref{eqn:meritQE}-\ref{eqn:meritTTS},
where the reference values for detection efficiency ($\epsilon$),
dark noise rate ($R_{DN}$) and transit time spread ($\sigma_{TTS}$)
are 27\%, 20 kHz and 1 ns, respectively.
Eq.~\ref{eqn:meritQE}-\ref{eqn:meritTTS}
quantitatively describe how an improvement on PDE can be
counteracted by an increase on dark noise rate or transit time spread.
The right side of each formula is scaled to represent a unified
price factor. For instance, if the PDE of all PMTs
increases by 1\%, the merit would be canceled if the price per PMT
increases by $\sim$1.2k CNY, as indicated by Eq.~\ref{eqn:meritQE} and
Eq.~\ref{eqn:priceMerit} (see Sec.~\ref{sec:selection}).

\begin{equation}
\label{eqn:meritQE}
M_{DE} = (\varepsilon-27)\times 3.518
\end{equation}
\begin{equation}
\label{eqn:meritDN}
M_{DN} = (20-R_{DN})\times 0.118
\end{equation}
\begin{equation}
\label{eqn:meritTTS}
M_{TTS} = -\log(\sigma_{TTS})
\end{equation}

Higher radioactivity of the PMT's glass would result in more
accidental background in the fiducial volume.
To keep the performance, we scale the liquid scintillator quantity to keep
the same signal-to-background ratio in the fiducial volume.
We convert the increment of liquid scintillator into a merit in Eq.~\ref{eqn:meritRad},
where $R_{\rm U}$, $R_{\rm Th}$ and $R_{\rm K}$ refer to the $^{238}$U, $^{232}$Th
and $^{40}$K contaminations in the PMT glass, respectively.
The measured values of low radioactive Schott glass,
22 ppb for $^{238}$U, 20 ppb for $^{238}$Th and
3.54 ppb for $^{40}$K~\cite{schott}, were taken as the reference.

\begin{equation}
\label{eqn:meritRad}
M_{Rad} = -\log\left[\dfrac{1}{3}\times \left( \dfrac{R_{Th}}{20} + \dfrac{R_U}{22} + \dfrac{R_K}{3.54} \right)\times 1.663\right]
\end{equation}

As for pre-pulses and after-pulses, a conservative
assumption was made that, the pre-pulses and after-pulses which
fall into the signal window would entirely contribute to the energy
scale uncertainty. The measured time distributions of pre-pulses and
after-pulses for each type of PMTs were used to calculate this effect.
Furthermore, the dependence of the neutrino-mass-ordering sensitivity
on the residual energy non-linearity is studied in Ref.~\cite{Li:2013zyd}.
Using those data, the merit for pre-pulses and after-pulses is
parameterized as Eq.~\ref{eqn:meritPA}, where $pp$ is the pre-pulse
probability and $ap$ is the after-pulse probability in percentage unit.
The after-pulses have much less impact because of its wider time
distribution.
\begin{equation}
\label{eqn:meritPA}
M_{PA} = -0.693\cdot(pp + 0.03\cdot ap)^{1.56}
\end{equation}

\begin{figure}[!htb]
\centering
\includegraphics[width=0.65\textwidth]{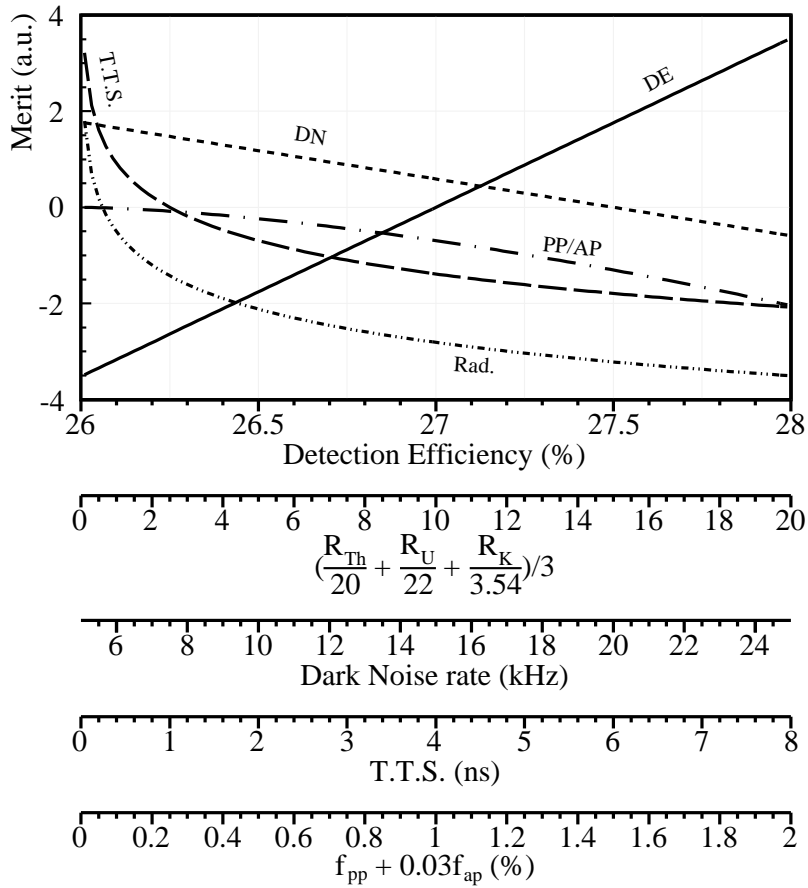}
\caption{The PMT performance merit for physics as functions of different PMT characteristics, including the detection efficiency ($M_{DE}$),
dark noise rate ($M_{DN}$), glass radioactivity ($M_{Rad}$),
pre-pulsing and after-pulsing ($M_{PA}$), and transit time spread ($M_{TTS}$). } \label{fig:merit}
\end{figure}

The merit of each PMT characteristic discussed above is shown in
Fig.~\ref{fig:merit}.
The total merit of the PMT performance, denoted as $M^{phys}$,
quantifies the overall impact of key parameters on physics goals:
\begin{equation}
\label{eqn:meritAll}
M^{phys} = M_{DE} + M_{DN} + M_{TTS} + M_{Rad} + M_{PA}
\end{equation}

\subsubsection{Selection strategy}
\label{sec:selection}

Most experiments have only one vendor for one type of PMTs,
as shown in Table.~\ref{tab:exp}.
However, one PMT type is more desirable in some
characteristics but less satisfactory in other parameters.
In addition, multiple vendors may reduce the risk if one of the
vendors is not able to deliver.
Thus, certain combination of different types of PMTs
could be the best choice, but this causes complications
to the selection strategy.
The right selection strategy is extended from Fig.~\ref{fig:diag}.

Multiple vendors participated in the bidding of JUNO's 20k PMTs.
In order to evaluate possible combinations of different
vendors with different award fractions,
the total 20k PMTs were divided into twenty packages.
Each vendor has the possibility to earn from 0 to 20 packages,
and each vendor is required to provide quotation for each possibility.
It is expected that the price per PMT is lower if the vendor earns
more packages. In this case, if three vendors participate,
231 combinations should be considered, while if two vendors participate,
only 11 combinations should be considered.

For each combination, we denotes that the $i$-th vendor earns $N_i$ packages,
and the corresponding quotation of price per PMT is $M_i$,
which is in 10k CNY unit.
Then, we assign a price factor for the $i$-th vendor:
\begin{equation}
\label{eqn:priceMerit}
M_i^{price} = 30\times(2.5 - M_i),
\end{equation}
Furthermore, a safety factor $S_i = 1 - k_S\cdot \eta_i$ is assigned to
the $i$-th vendor, where $k_S$ is arbitrarily set to 0.15.
The motivation is simple, the larger award fraction one vendor
wins, the higher risk we takes.
According to the PMT specifications that the $i$-th vendor guarantees,
the merit of physics $M^{phys}_i$ is given by Eq.~\ref{eqn:meritAll}.
Finally, a selection committee gives overall justification
to each vendor, denoted as $M_i^{committee}$ for the $i$-th vendor,
with the main considerations on technology reliability, production schedule,
and service commitments, etc. The highest and lowest marks among the
individual committee members were removed and the average of the rest was
taken as $M_i^{committee}$. In the end, the total merit of a specific
combination is a summation over the vendors, as shown below:
\begin{equation}
\label{eqn:finalScore}
S = \sum (M_i^{phys} + M_i^{price} + M_i^{committee})\cdot S_i \cdot N_i/20,
\end{equation}
where the importance of $M_i^{phys}$, $M_i^{price}$ and $M_i^{committee}$
are designed to be 40\%, 30\% and 30\%, respectively.
The final decision on PMT choice is the combination with
the maximum total merit $S$.

\subsubsection{Result}

The public bidding for JUNO PMTs was opened in Dec, 2015
with two options: 20k 20-inch PMTs as the baseline,
or 130k 8-inch PMTs as the alternative which has
roughly equivalent total photocathode coverage.
The bidders were required to provide the guaranteed PMT
characteristics and the quotation for at least one of the two options.
Three companies provided quotations for the 8-inch PMT option, but
the total price was too high, thus this option was abandoned.
Two of the three companies, NNVT and Hamamatsu, provided quotations
with 20-inch MCP-PMT (GDG-6201) and dynode-PMT (R12860) products, respectively.
Besides the main characteristics shown in Table.~\ref{tab:pmtPar},
other characteristics such as the mechanical strength under water,
the long-term stability and the performance under Earth's magnetic field
all satisfy the minimum specifications of JUNO.
With the guaranteed PMT characteristics and quotations,
the quantitative approach described in this section was used
to compare different combinations of MCP-PMTs and dynode-PMTs.
The combination of 15,000 20-inch MCP-PMTs and 5,000
20-inch dynode-PMTs gave the best total merit,
thus it was the final choice of JUNO PMTs.
Accordingly to the merit of physics performance (see Sec.~\ref{sec:physMerit}),
having a quarter of PMTs to be dynode type is mainly due to
its good transit time spread which helps the vertex reconstruction.

\section{Summary}
\label{sec:summary}

Large area PMTs are likely the first choice
of photon sensors for large liquid scintillator or water neutrino
experiments which require high photocathode coverage.
Besides photon detection efficiency, other characteristics such as
dark noise rate, transit time spread, radioactive background of glasses,
peak-to-valley ratio, etc, will affect the photon detection and
hence affect the physics goals.
It is theoretically and practically important to build a detector
with maximum physics potential and minimum cost.
This needs a science-driven approach to quantify the impact
of all PMT parameters,
including cost and risk, to the physics goals, and a selection strategy
to make the decision.
The quantitative approach described in this paper,
which was successfully used for JUNO 20-inch PMTs selection,
would set up a reference for future projects.
After all, optimizing the selection of PMTs to build state-of-art
neutrino observatories will enhance future scientific outputs.

\vspace{12pt}
\section*{ACKNOWLEDGMENTS}
This work was supported by the Strategic Priority Research Program of the Chinese Academy of Sciences, Grant No. XDA10010100; the CAS Center for Excellence in Particle Physics (CCEPP) (for all authors).

\end{document}